\documentclass{JINST}
\usepackage{lineno}
\usepackage[percent]{overpic}
\usepackage{amsmath}
\usepackage{amssymb}
\newcommand{\DBD}{0$\nu$DBD}

\newcommand{\TEO}{$\mathrm{TeO}_2$}
\newcommand{\ZNMO}{$\mathrm{ZnMoO}_4$}

\newcommand{\Cuore}{CUORE}

\providecommand*{\un}[1]{\ensuremath{\mathrm{~#1}}}

\title{Performances of a large mass ZnSe bolometer to search for rare events} 

\author{
J.W.~Beeman$^a$, 
F.~Bellini$^{b,c}$, 
L.~Cardani$^{b,c}$\thanks{Corresponding author.}, 
N.~Casali$^{d,e}$,
I.~Dafinei$^c$, 
S.~Di~Domizio$^h$, 
F.~Ferroni$^{b,c}$, 
L.~Gironi$^{f,g}$, 
A.~Giuliani$^i$,
S.~Nagorny$^e$, 
F.~Orio$^c$, 
L.~Pattavina$^e$, 
G.~Pessina$^g$, 
G.~Piperno$^{b,c}$, 
S.~Pirro$^g$, 
E.~Previtali$^g$
C.~Rusconi$^g$, 
C.~Tomei$^c$, 
M.~Vignati$^c$\\
\llap{$^a$}Lawrence Berkeley National Laboratory , Berkeley, California 94720, USA\\
\llap{$^b$}Dipartimento di Fisica - Universit\`{a} di Roma La Sapienza, I-00185 Roma - Italy\\
\llap{$^c$}INFN - Sezione di Roma, I-00185 Roma - Italy\\
\llap{$^d$}Dipartimento di Scienze Fisiche e Chimiche - Universit\`{a} degli studi dell'Aquila, I-67100 Coppito (AQ) - Italy\\
\llap{$^e$}INFN - Laboratori Nazionali del Gran Sasso, I-67010 Assergi (AQ) - Italy\\
\llap{$^f$}Dipartimento di Fisica - Universit\`{a} di Milano Bicocca, I-20126 Milano - Italy\\
\llap{$^g$}INFN - Sezione di Milano Bicocca, I-20126 Milano - Italy\\
\llap{$^h$}INFN - Sezione di Genova, I-16146 Genova - Italy\\
\llap{$^i$}CSNSM, Centre de Spectrom\'etrie Nucl\'eaire et de Spectrom\'etrie de Masse, B\^atiment 108, Campus d'Orsay, 91405 Orsay, France
\\
E-mail:\email{laura.cardani@roma1.infn.it}
}

\abstract
{
Scintillating bolometers of ZnSe are the baseline choice of the LUCIFER experiment, whose aim is to observe
the neutrinoless double beta decay of $^{82}$Se.
The independent read-out of the heat and scintillation signals
allows to identify and reject $\alpha$ particle interactions, the dominant background source for bolometric detectors.
In this paper we report the performances of a ZnSe crystal operated within the LUCIFER R\&D.
We measured the scintillation yield, the energy resolution and the background in the energy region where the signal from \DBD \ decay of $^{82}$Se is expected
with an exposure of $9.4\un{kg\cdot days}$.
With a newly developed analysis algorithm we improved the rejection of $\alpha$ events,
and we estimated the increase in energy resolution obtained by the combination of the heat and light signals.
For the first time we measured the light emitted by nuclear recoils, and found it to be compatible with zero.
We conclude that the discrimination of nuclear recoils from $\beta/\gamma$ interactions in the WIMPs energy region
is possible, but low-noise  light detectors are needed.
}


\begin{document}

\section{Introduction} 

Bolometers are solid state detectors in which the energy release coming from particle interactions
is converted to heat and measured via their rise in temperature. 
They can provide excellent energy resolution and low background, and are used
in particle physics experiments searching for rare processes, such as
neutrinoless double beta decay (\DBD) and Dark Matter interactions.

In the last years, an extensive R$\&$D allowed to increase the mass of bolometric detectors
and, at the same time, to understand and reduce the sources of background.
The largest bolometric detector operated up to now, Cuoricino, was made of about 40.7 kg of TeO$_2$ and
took data from 2003 to 2008, demonstrating the potential of this technique~\cite{Ndrello3,Andreotti:2010vj}.
The evolution of Cuoricino, \Cuore, will search for the \DBD \ of $^{130}$Te~\cite{Ardito:2005ar, ACryo} using an array of 988 \TEO\ bolometers
of 750\un{g} each. Operated at a temperature of about 10\un{mK}, these
detectors provide an energy resolution of a few keV over their energy
range, extending from a few keV up to several MeV.  The measured resolution
at the Q-value of the decay ($Q=2527\un{keV}$~\cite{Rahaman:2011zz}) is about 5\un{keV\,FWHM}; 
together with the low background and the large mass of the experiment, 
this will provide a 1$\sigma$ sensitivity to the \DBD \ of $^{130}$Te of the order of 10$^{26}$ years.

To further increase the sensitivity, an intense R\&D is being pursued to
lower the background in the \DBD\ region. The main source of background
is due to $\alpha$ particles, coming from radioactive contaminations
of the materials facing the bolometers~\cite{Clemenza:2011zz}. A way to discriminate this
background  is  to use a scintillating bolometer~\cite{Pirro:2005ar}.
In such a device the simultaneous and independent read-out of the
heat and the scintillation light permits to discriminate events due to
$\beta / \gamma$, neutron and $\alpha$ interactions thanks to their different
scintillation properties.  Unfortunately \TEO\  crystals do not scintillate 
and different compounds are being studied. 
Among these are CdWO$_4$~\cite{Arnaboldi:2010tt}
(\DBD\ candidate $^{116}$Cd, $Q=2814\un{keV}$~\cite{Rahaman:2011zz}),
ZnSe ~\cite{Arnaboldi:2010jx} ($^{82}$Se, $Q=2997\un{keV}$~\cite{Lincoln:2012fq}),
and \ZNMO\ ~\cite{Gironi:2010hs,Beeman:2011bg,Beeman:2012gg,ZnMoO4_1} ($^{100}$Mo, $Q=3034\un{keV}$~\cite{Rahaman2008111}).

In this paper we investigate the performances of a ZnSe scintillating crystal for the \DBD\ search in terms of
energy resolution, capability of discriminating $\alpha$ particles and
internal radioactive contaminations. We also study the light emitted by nuclear recoils
and $\beta/\gamma$ interactions at 100\un{keV}, discussing the possibility
of using ZnSe bolometers as Dark Matter detectors.

\section{Experimental setup}
The data here presented come from a series of runs performed at Laboratori Nazionali del Gran Sasso (LNGS) in Italy, inside the CUORE R\&D facility.
The bolometer under study is a 431\un{g} ZnSe crystal with cylindrical shape
(height 44.3\un{mm} and diameter 48.5\un{mm}).  The synthesis of the
ZnSe powder, including purification and formatting for crystal growth~\cite{Ryzhikov2013111,Rudolph199585}, 
was made in ultra-clean fused quartz reactors at SmiLab Svitlovodsk (Ukraine).

To detect the scintillation light, many different light detectors (LD) were faced to the ZnSe crystal.
They consist of pure Germanium slabs of 50\un{mm} diameter and with variable thickness (300 - 600 \un{\mu m})
and they were operated as bolometers to obtain good performances at cryogenic temperatures~\cite{Pirro:2006ra}.
We covered a face of the LD with a thin layer of SiO$_2$ (60\un{nm}) and obtained an increase of the light absorption  by $\sim$ 16 - 20$\%$,
as already observed in Ref.~\cite{Beeman:2012cu}. 
Moreover, we measured an increase of  $\sim$ 25$\%$ in the collection efficiency by surrounding the crystal with a 3M VM2002 reflecting foil.

The temperature sensors of both the ZnSe and LD were Neutron Transmutation Doped (NTD) Germanium
thermistors~\cite{Itoh:1996}, coupled to the ZnSe and Ge surfaces by means
of epoxy glue spots.  For redundancy, the ZnSe crystal was equipped with
two sensors, in the following referred as ZnSe-L and ZnSe-R.
The detectors were held in a copper structure by Teflon (PTFE) supports and
thermally coupled to the mixing chamber of a dilution refrigerator which
kept the system at a temperature around $10\un{mK}$. 
To read the signals, the thermistors were biased with a constant current. The
voltage signals, amplified and filtered by means of an anti-aliasing
6-pole active Bessel filter (120 dB/decade), were fed into a NI PXI-6284
18-bits ADC operating at a sampling frequency of 2 kHz. The Bessel
cutoff was set at 120\un{Hz} for the LDs, and at 70\un{Hz} for the ZnSe.
Further details on the cryogenic facility and the electronic read-out can be
found in Refs.~\cite{Pirro:2006mu, Arnaboldi:2006mx,Arnaboldi:2004jj}.

The trigger was software generated on each bolometer. When it fired,
waveforms 5\un{s} long on the ZnSe and 250 ms long on the LD were digitized
and saved on disk. In addition, for every trigger occurring on the
ZnSe, the LD waveform was acquired irrespective of its trigger.  The off-line
analysis computes the pulse height as well as pulse shape
parameters based on the optimum filter algorithm~\cite{Gatti:1986cw,Radeka:1966}.  In addition,
the amplitude of the light signals is computed with an 
algorithm that allows to lower the energy threshold of the LD using the knowledge of the
signal time delay with respect to the ZnSe (see details in Ref.~\cite{Piperno:2011fp}).

\begin{figure}[tb]
\centering
\includegraphics[clip=true,width=0.45\textwidth]{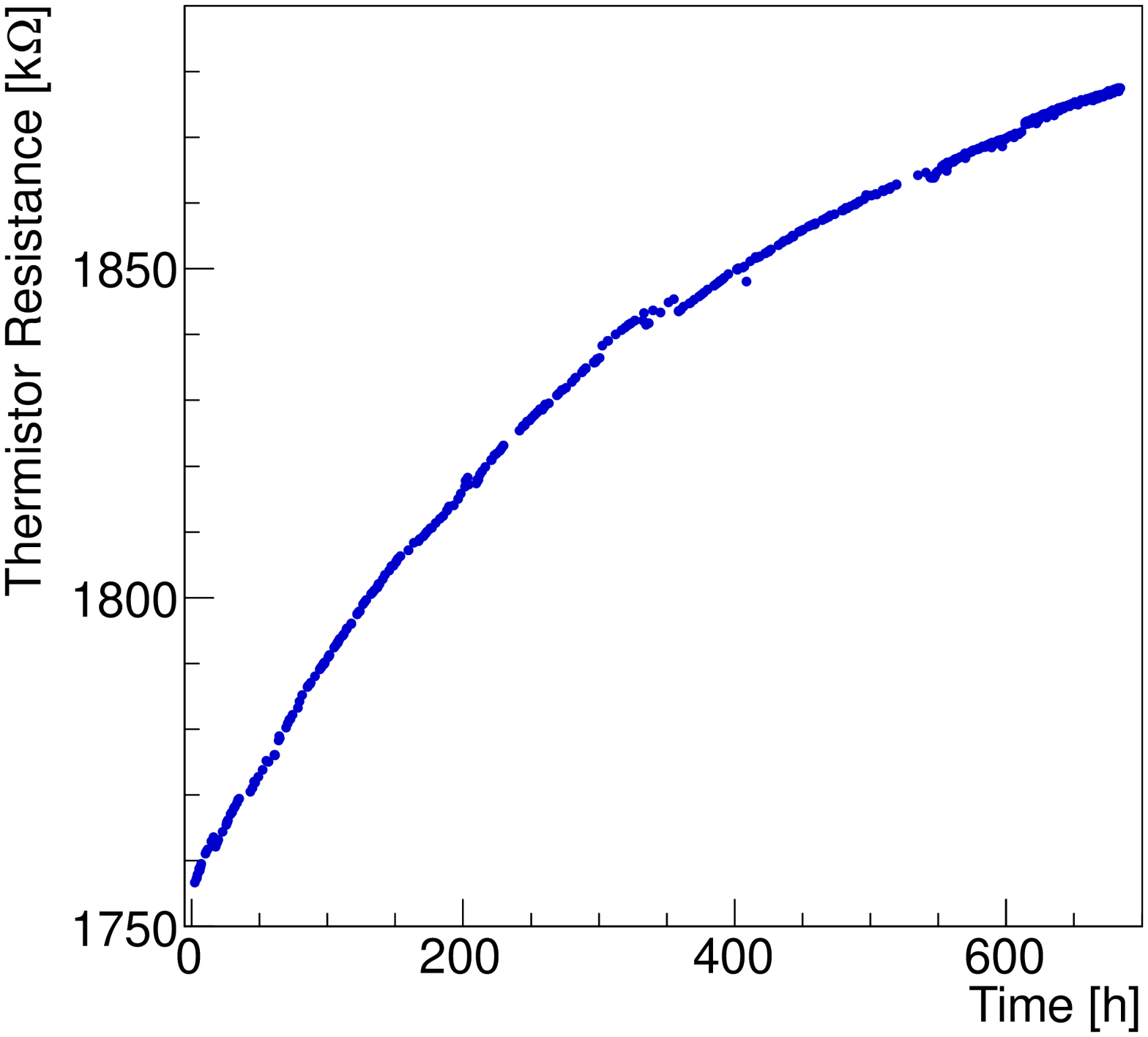}
\hfill
\includegraphics[clip=true,width=0.45\textwidth]{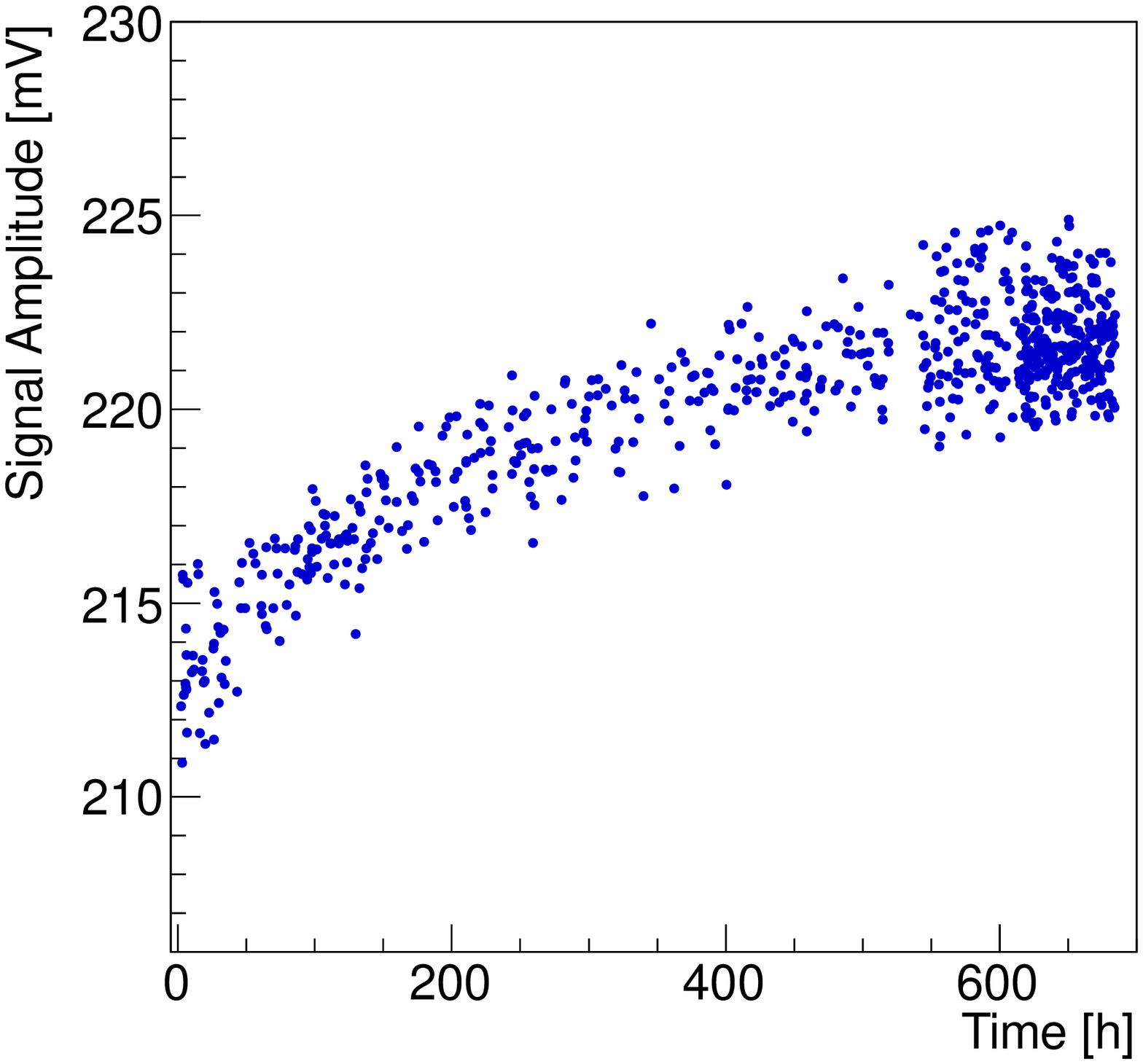}
\caption{Change in the thermistor response induced by the slow cooling of the ZnSe bolometer, whose temperature decreased of about 1$\%$ during the first 700 hours of measurement.
Left: increase of the thermistor resistance due to the cooling of the crystal. 
Right: amplitude of pulses with energy of about 1460~\un{keV} as a function of time.
In the last 150 hours a calibration measurement was performed, resulting in a higher rate.}
\label{fig:ZnSe.Drift}
\end{figure}

Compared to \TEO\ crystals, we
observed that the ZnSe  bolometer cooled very slowly, likely because
of a heat capacitance excess at low temperature.
Indeed, despite of the fact that the cryostat temperature was kept constant, 
the temperature drift of the ZnSe was large enough to influence the detector response still after 30 days of  data-taking (Fig.~\ref{fig:ZnSe.Drift}). 
The resistance of the thermistor, which has a steep dependence on the temperature, increased of about 6$\%$ during this data acquisition period,
resulting in a corresponding increase in the signal height.
This behavior was also observed in bolometers made of BGO crystals~\cite{Cardani:2012xq}.


The  calibration of the ZnSe pulse amplitude was performed by means of a $^{228}$Th
$\gamma$-source placed inside the cryostat external lead shield.
We observed that
the calibration function derived from $\gamma$ peaks was not 
applicable to energy deposits induced by $\alpha$ particles,
which were shifted by $+22\%$ with respect to their nominal energy.
For this reason, in the following we will refer to the energy estimated from the $\gamma$ calibration as ``Energy$_{ee}$''
($\gamma$s and electrons interact in the same way in the bolometer).
When the calibration is derived from $\alpha$ peaks 
produced by the internal contaminations, the energy units will be indicated as  ``Energy$_{\alpha}$''.\\
To evaluate the discrimination power between $\beta/\gamma$ and
$\alpha$ events in the energy region of interest (around 2997 keV~\cite{Lincoln:2012fq}),
an $\alpha$-source was permanently placed close to the ZnSe crystal.
The source consisted in an
Uranium solution, 
covered with a thin mylar foil to absorb part of the $\alpha$s
energy and to produce a continuum spectrum in the range 1-4\un{MeV}.
Finally, during a calibration run, an AmBe neutron source was placed
close to the detector in order to produce high energy $\gamma$s.

The LD is calibrated with a $^{55}$Fe source permanently faced
to the LD surface opposite to the ZnSe.  The source emits two
X-rays at 5.9 and 6.5~keV.

The main features of the detectors are summarized in
Table~\ref{Table:parameters_crystals}.
The rise and decay times of the pulses are defined as the time difference between the 90\% and the 10\% of the leading edge, 
and the time difference between the 30\% and 90\% of the trailing edge, respectively. 
The intrinsic energy resolution of the detector ($\sigma_{baseline}$) is
estimated from the fluctuations of the detector baseline after the optimum filter application. Because of the better energy resolution, in the following we restrict our analysis to ZnSe-R.
\begin{table}[hbtp]
\centering
\caption{Parameters of the bolometers. Amplitude of the signal before  amplification  ($A_S$),  intrinsic energy resolution after the application of the optimum filter  ($\sigma_{baseline}$), rise ($\tau_r$) and decay ($\tau_d$) times of the pulses. ZnSe-R is the thermistor chosen for the data analysis. }
\begin{tabular}{lcccc}
\hline
                       & $A_S$                                             &$\sigma_{baseline}$     &$\tau_{r}$              &$\tau_{d}$\\
                       &[$\mu$V/MeV]                                 &[keV RMS]            &[ms]                        &[ms]   \\
\hline
ZnSe-L           & 10                                                  & 5.2              &5.6 $\pm$ 0.3       &30 $\pm$ 4    \\
\hline
ZnSe-R          & 50                                                  & 2.4                 &5.7 $\pm$ 0.2       &23 $\pm$ 2\\
\hline
LD      	      &980                                                 &0.069                           &2.8$\pm$ 0.1        &9.4 $\pm$ 0.5    \\
\hline
\end{tabular}
\label{Table:parameters_crystals}
\end{table}

\section{Scintillation}
\label{sec: scintillation}

\begin{figure}[tb]
\centering
\includegraphics[clip=true,width=0.7\textwidth]{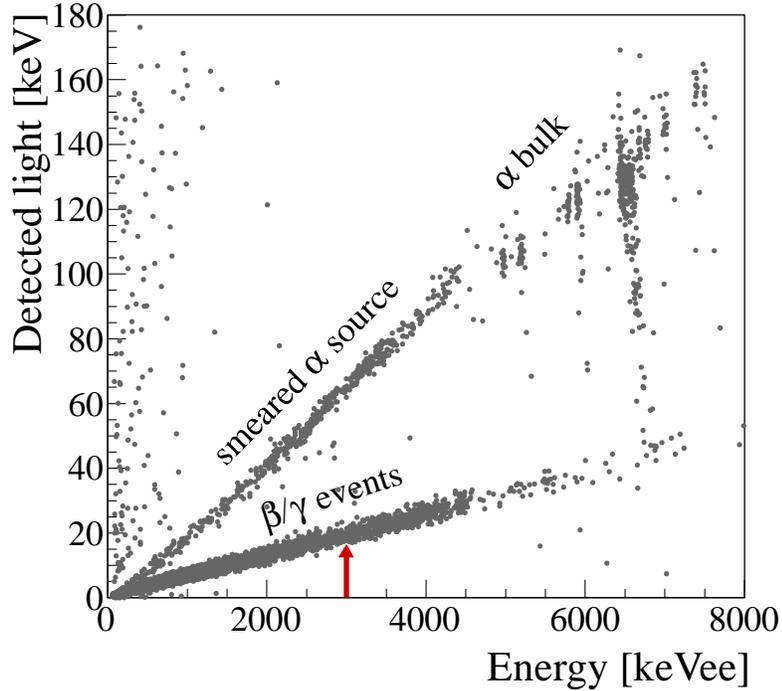}
\hfill
\caption{Detected Light vs Energy measured by the ZnSe in a calibration run. The energy axis has been calibrated using the most intense $\gamma$ peaks.
The ZnSe crystal is faced to a smeared $\alpha$ source (upper band) and to an AmBe neutron source (high energy $\gamma$s in the lower band). 
The events characterized by low energy (below $\sim$ 1 MeV) and large detected light are particles that interacted in both the ZnSe and LD.
The arrow points to the $^{82}$Se Q-value (2997\un{keV}).}
\label{fig:light vs heat calibration} 
\end{figure}

The measured light as a function of the energy deposited in the ZnSe bolometer
is shown in Fig.~\ref{fig:light vs heat calibration}. Two
regions can be clearly distinguished: the upper band, populated by $\alpha$
events provided by the smeared $\alpha$ source (continuum below $\sim$\
4.5 MeV$_{ee}$) and by internal and surface $\alpha$ contaminations, and the
lower band, populated by $\gamma$s produced by the neutron source.
The Light Yield of $\beta/\gamma$ events (LY$_{\beta/\gamma}$), defined as the
amount of detected light per particle energy, does not depend on the energy.
Fitting the $\beta/\gamma$ band with a first order polynomial function,
we obtain LY$_{\beta/\gamma}$ = 6.416 $\pm$ 0.008 keV/MeV.  The Light Yield
of $\alpha$s (LY$_{\alpha}$) is larger than LY$_{\beta/\gamma}$, unlike in
other known scintillating crystals~\cite{Tretyak:2009sr}.
This behavior, already observed in other ZnSe crystals operated at
cryogenic temperatures, is not yet fully understood.  Nevertheless, as
it will be shown in Sec.~\ref{sec:psa}, this does not limit the discrimination capability.

The continuum produced by the smeared $\alpha$ source was fitted
with a first order polynomial function, resulting in
LY$_{\alpha}^{smeared}$ = (29.70 $\pm$ 0.17) keV/MeV.  The fit revealed
the presence of an energy threshold for the scintillation,
which is E$_{thresh}^{smeared}$ = (230 $\pm$ 12) keV$_{\alpha}$.  At larger energies,
where only $\alpha$ peaks due to bulk  contaminations are present,
we obtained  LY$_{\alpha}^{bulk}$ = (26.62 $\pm$ 0.86) keV/MeV
and E$_{thresh}^{bulk}$ = (180 $\pm$ 140) keV$_{\alpha}$.  The large error on the
threshold is due to the large distance of the peaks from the origin, and
does not allow a trustworthy comparison with the smeared source. However
it is clear that LY$_{\alpha}^{bulk}$ is lower than LY$_{\alpha}^{smeared}$.

Understanding this discrepancy is not trivial. 
Indeed, for a given light detector and experimental set-up, the light yield includes not only the scintillation process that produces light, 
but also the light transport from the luminescent centre to the detector.
Even if we are not able to disentangle these two processes, 
we performed some tests to investigate the effects  on the
discrepancy between LY$_{\alpha}^{bulk}$ and LY$_{\alpha}^{smeared}$ induced by the crystal self-absorption, 
by the reflection on the mylar foil and by the energy dependence.
In detail:
\begin{itemize}
\item we checked whether the lower LY of bulk events could be due to the ZnSe self-absorption. 
This test was performed by placing the smeared source and the LD on opposite sides of the ZnSe, 
in such a way that the light emitted by $\alpha$s impinging on the surface had to travel across the entire crystal before reaching the LD.
The results of this test are the ones already reported above and show that   LY$_{\alpha}^{smeared} > LY_{\alpha}^{bulk}$ 
even if we expect the maximum re-absorption of the scintillation light emitted by the smeared $\alpha$s.
\item we investigated a possible energy dependence of the LY by facing a $^{224}$Ra source emitting high energy $\alpha$s  to the lateral surface of the ZnSe crystal,
so that we could reproduce the same $\alpha$ particles from the bulk contamination and compare external and internal sources at the same energy.
We obtained LY$_{\alpha}^{external}$ = (29.6 $\pm$ 0.1) keV/MeV from 4 to 6 MeV. 
This value is compatible with LY$_{\alpha}^{smeared}$ pointing to the hypothesis that the larger LY is not due to energy-dependent effects.
We underline that this test was performed without reflecting foil, to make sure that the larger LY of external $\alpha$s could not be attributed to 
the reflection on the mylar sheet.
\end{itemize}
The experimental data show that in every test  the light yield from bulk interactions is lower than the one produced by external events.
Further measurements are needed to deepen the understanding of this behavior, 
that could be attributed  to an effectively higher light production 
(the bulk of a real crystal is different in many aspects from the thin layer close to the surface),
or to the different light collection for surface/bulk events.
Indeed, it is well known that the light collection is particularly non uniform when dealing with cylindrical crystals~\cite{Danevich:2002rg}.\\

To evaluate the discrimination capability at the \DBD\ energy, 
we performed Gaussian fits to the light emitted in $\alpha$ and $\beta/\gamma$ interactions,  
excluding the outliers due to $\alpha$ interactions in which a leakage of light is detected.
From the fits we derived the mean value ($\mu$) and the standard deviation  ($\sigma$) of the light
for both the $\alpha$ and $\beta/\gamma$ interactions. 
Since the discrimination capability increases with energy (as one can see in Fig. \ref{fig:light vs heat calibration}),
we calculated  $\mu$ and $\sigma$ in several energy intervals and fitted the energy dependence of  $\mu$(E) and $\sigma$(E)
with polynomial functions.\\
We defined the Discrimination Potential as a function of the energy as:
\begin{equation}
DP(E) = \frac{\left|\mu_{\alpha}(E)-\mu_{\beta\gamma}(E)\right|}{\sqrt{\sigma_\alpha^2(E)+\sigma_{\beta\gamma}^2(E)}}
\end{equation}
and found $DP=17$ at 2997\un{keV}. It has to be remarked that this is just and indication
of the capability of rejecting the $\alpha$ background. 
As it can be noticed from Fig.~\ref{fig:light vs heat calibration}, there is a considerable number of $\alpha$s in which a large amount of light is lost. 
This behavior, likely due to surface effects, is particularly evident looking at the peak of $^{210}$Po ($\sim$ 6.5 MeV$_{ee}$), 
a common contaminant found on the ZnSe surface.
The loss of light from $\alpha$ particles constitutes a non-negligible and hard to estimate background to $\beta/\gamma$ events. 
As it will be shown in Sec.~\ref{sec:psa}, the pulse shape of the
light signal carries information on the particle type, irrespective
of the amount of light collected, allowing a safe identification of
all $\alpha$ events.

\begin{figure}[tb]
\centering
\includegraphics[clip=true,width=0.45\textwidth]{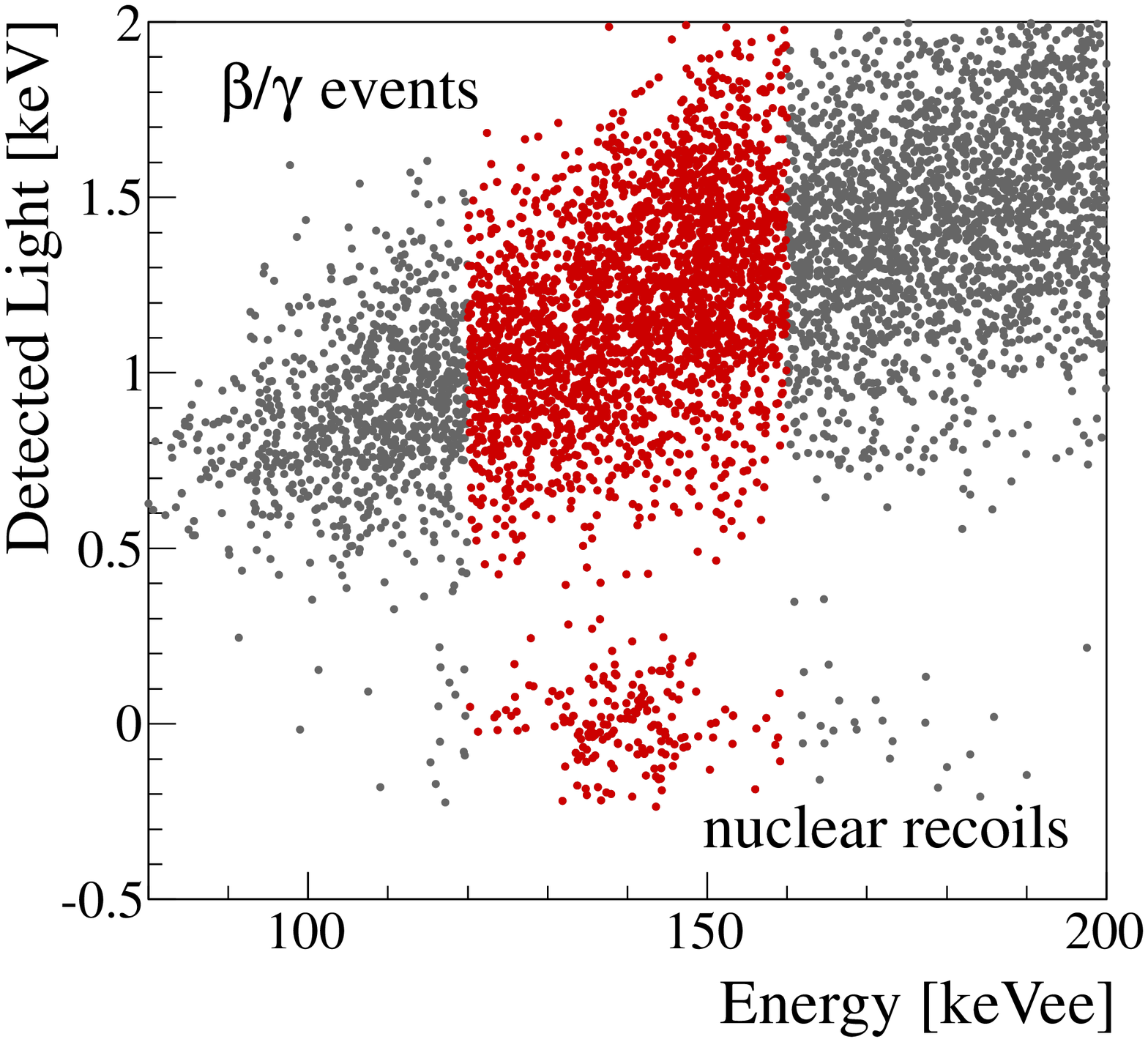}
\hfill
\includegraphics[clip=true,width=0.45\textwidth]{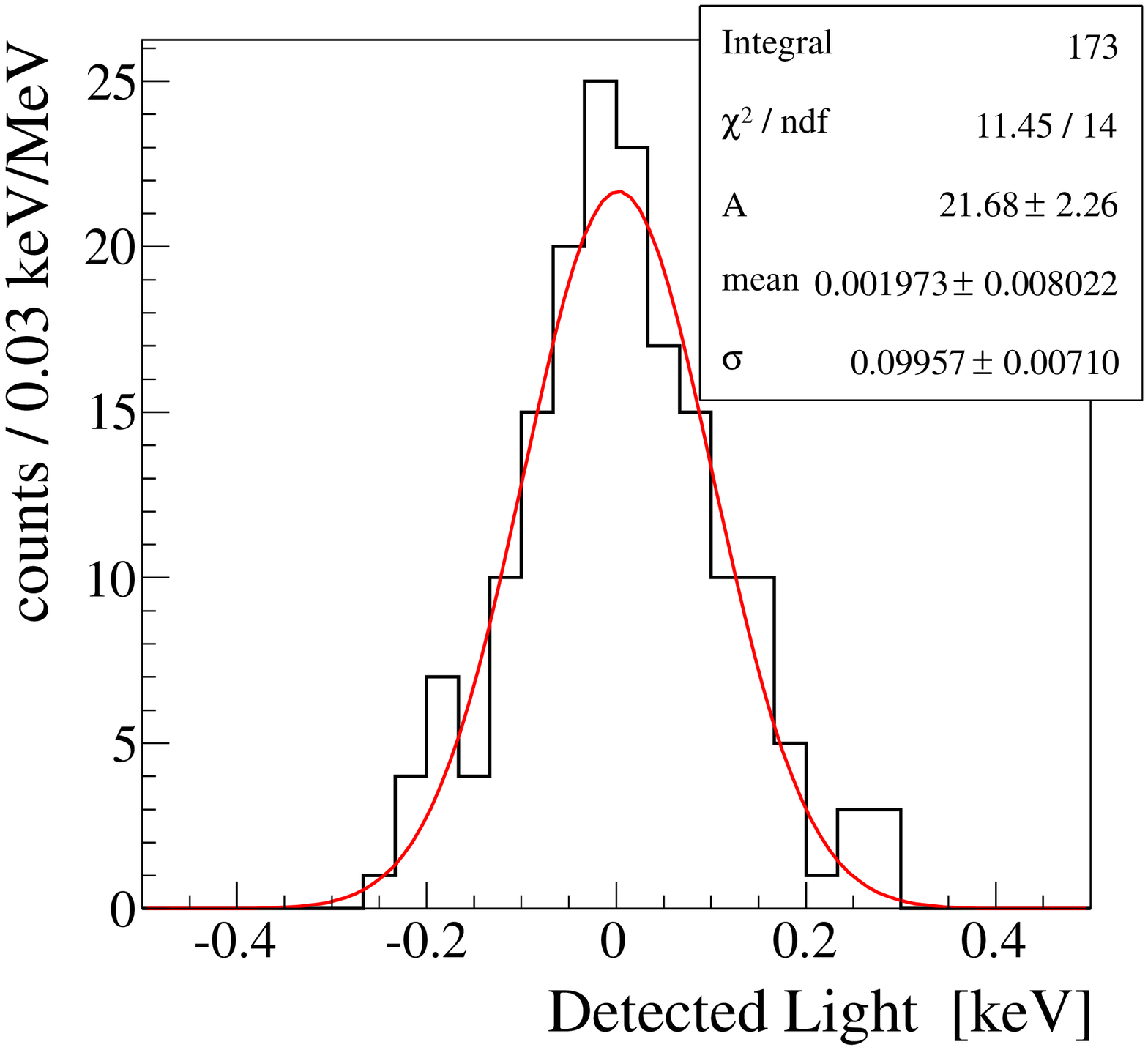}
\caption{Left: Energy region around the nuclear recoils from the $^{210}$Po decay.  
The detected energy of the recoils, using the $\beta/\gamma$ calibration function, is $35\%$ larger than the nominal value (103\un{keV}, see discussion in the text).
Right: Distribution of the light emitted from the nuclear recoils, selecting the events marked in red in the left figure and applying the cut Detected Light~<~0.3\un{keV} to remove the $\beta/\gamma$ background.}
\label{fig:recoils}
\end{figure}

We finally studied the LY at very low energies,
which is of particular interest for experiments aiming at the detection
of Dark Matter interactions.  These experiments need to disentangle the
signal produced by nuclear recoils below 30 keV (following the hypothesis that Dark Matter is made of WIMPs~\cite{Steigman:1984ac,Goodman:1984dc}) from the background induced by $\beta/\gamma$s.
We analyzed the recoils following the $\alpha$ decay of $^{210}$Po.
Since this isotope is deposited on the surface, the $\alpha$ particle can escape
without releasing energy, while the nuclear recoil is absorbed in the crystal.
We observe events centered at $139.6 \pm 0.6\un{keV_{ee}}$ in the ZnSe, i.e. 35\% more energy than the nominal value (103\un{keV}).
The accuracy of the calibration function has been checked down to 511 keV, where it shows a deviation
less than 1$\%$. The extrapolation at lower energies is expected to maintain or reduce this deviation.
From the fit in Fig.~\ref{fig:recoils}, we evaluate the light emitted by nuclear recoils as < 14 eV at 90$\%$ C.L., corresponding to 
LY$_{nr}$ < 0.140\un{keV/MeV} at 90$\%$ C.L. 
We evaluate the light emitted in the range $10-30\un{keV_{nr}}$ as $<1-4\un{eV}$ at 90\%~C.L.
for nuclear recoils, and $90-260\un{eV}$ for $\beta/\gamma$s.

In this test the energy threshold was not optimized and set at 70\un{keV_{ee}} ($\sim50\un{keV_{nr}}$, assuming the $35\%$ miscalibration), while to be competitive
with present experiments it should be below 10\un{keV_{nr}}. 
Given the baseline fluctuation ($2.4\un{keV_{ee}} = 1.7\un{keV_{nr}}$ ) the required threshold could be reached
with a trigger based on the optimum filter~\cite{DiDomizio:2010ph}.
However, to obtain a DP between $\beta/\gamma$s and nuclear recoils at least larger than $3$,
light detectors with baseline noise less than $20\un{eV~RMS}$ are needed.
The light detectors currently being used are far from this value (see Tab.~\ref{Table:parameters_crystals}) 
and obtaining resolutions better than 70\un{eV~RMS} does not seem to be achievable with the present technology.
To search for Dark Matter interactions in ZnSe bolometers, new light detection technologies must be introduced.

\section{Energy Resolution}

\begin{figure}[tb]
\centering
\includegraphics[clip=true,width=0.45\textwidth]{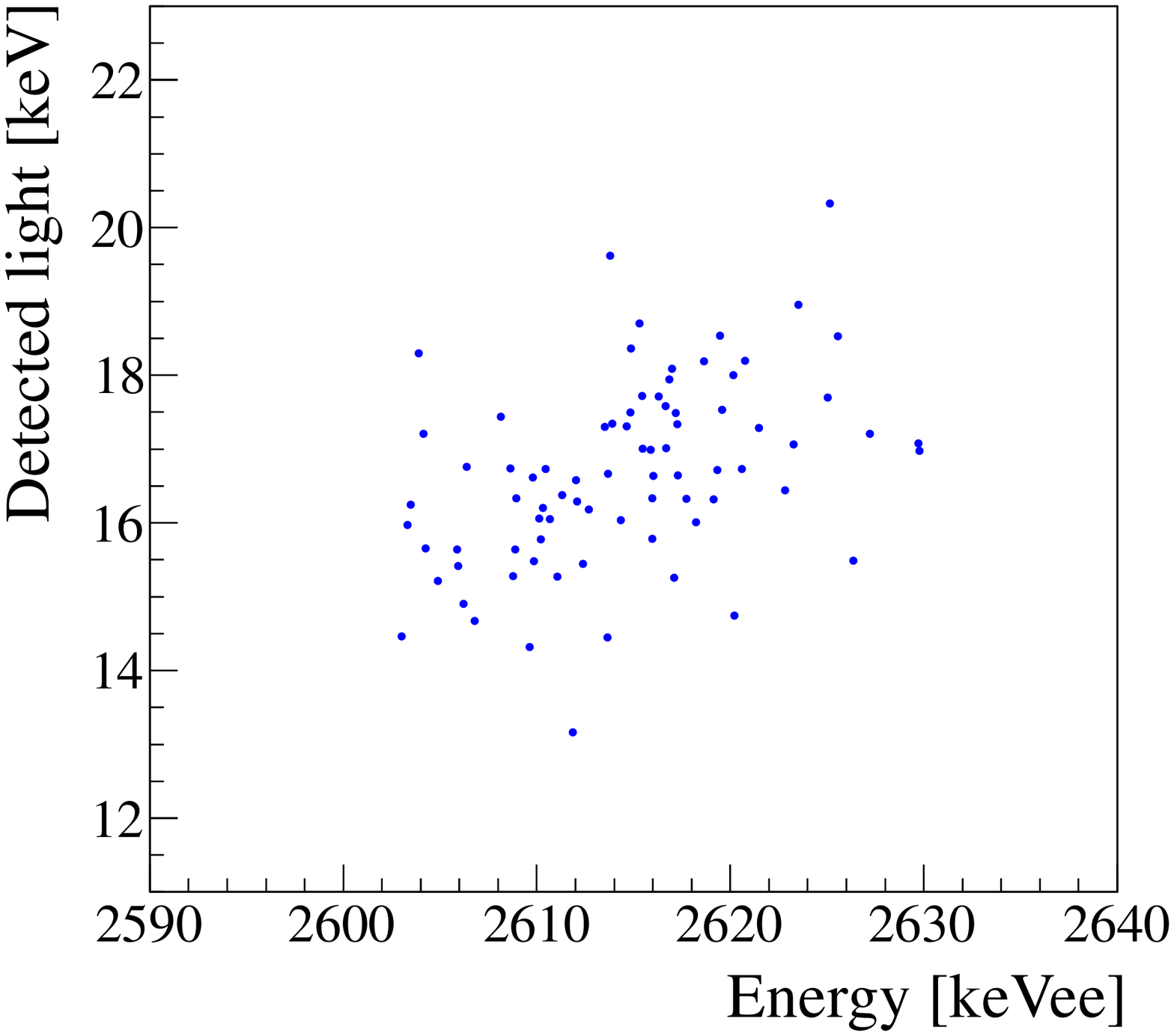}
\includegraphics[clip=true,width=0.45\textwidth]{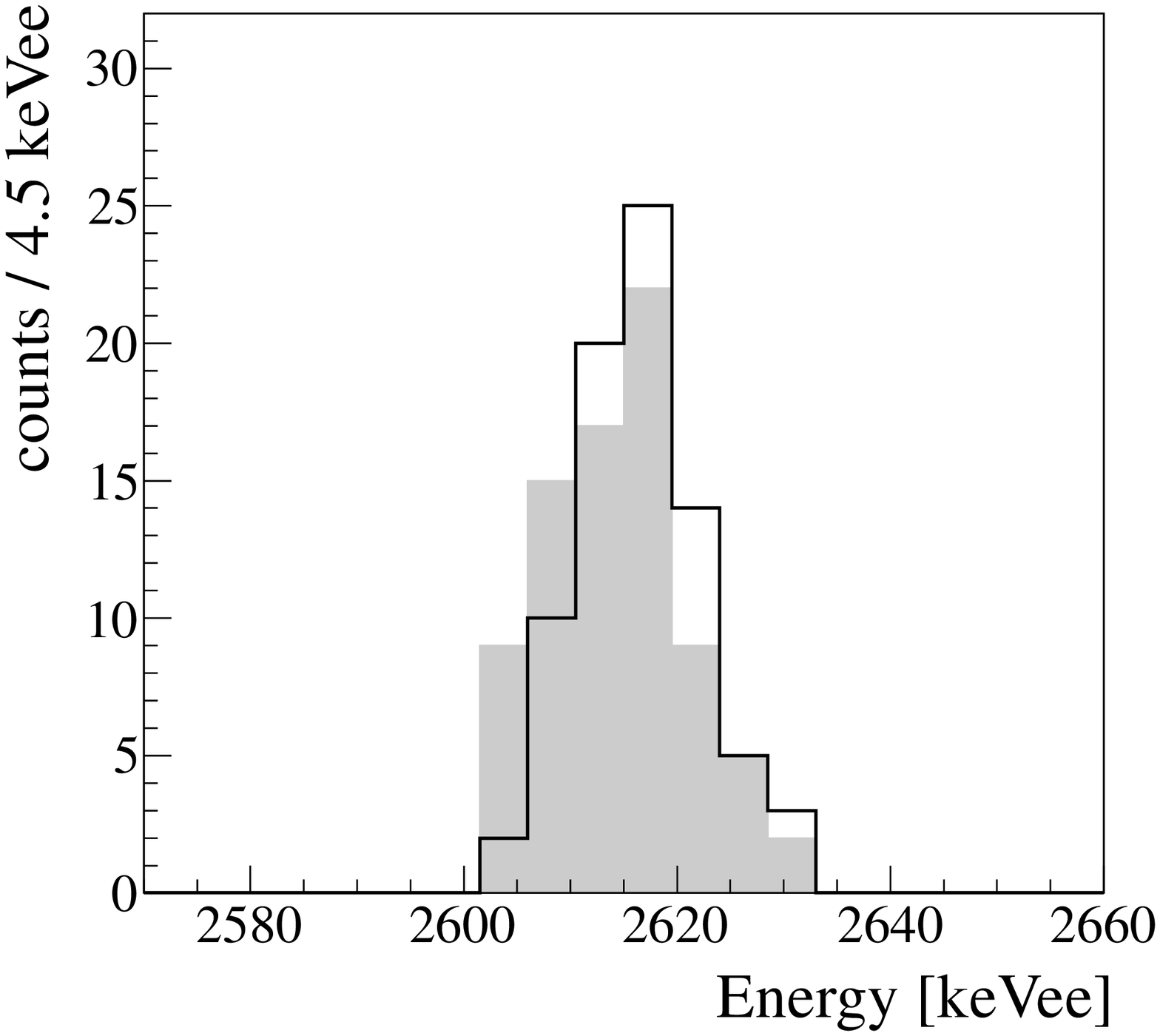}
\hfill
\caption{Left: Correlation between the detected light and the energy released in the ZnSe for 2615\un{keV} $\gamma$ events. 
Right: Energy distribution in the ZnSe before (gray) and after (black line) the combination with the light signal for the same events.}
\label{fig:decorrelation} 
\end{figure}

The energy resolution on the \DBD\ signal is estimated from $\gamma$ lines, which produce the same bolometric
and scintillation response of $\beta$ particles. The energy resolution  is found to be  $13.4\pm1.0~\un{keV~FWHM}$ at 1461\un{keV} ($^{40}$K contamination in the cryostat) 
and  $16.3 \pm 1.5\un{keV~FWHM}$ at 2615 keV ($^{208}$Tl from calibration). These values are worse than the baseline resolution, which is  $5.6\un{keV~FWHM}$ (see Tab.~\ref{Table:parameters_crystals}). 
Looking at the detected light, we observe that there is a positive correlation with the energy measured in the ZnSe (Fig.~\ref{fig:decorrelation} left). 
This is in contrast with other scintillating crystals, where the correlation is negative as expected from the conservation of energy ~\cite{Arnaboldi:2010tt}.
Presently we are unable to explain this behavior, however we can take advantage from the correlation to improve the energy resolution.
 
The light $L$ and energy $E$ in the ZnSe  are two correlated estimates of the same observable, i.e. the particle energy. 
Defining $L'=L/LY_{\beta/\gamma}$,  we write the combined energy variable as:
\begin{equation}\label{eq:goremagari}
E^{comb} = (1-w)\, E + w\, L'\;. 
\end{equation}
The weight $w$ that minimizes the variance of $E^{comb}$ is:
\begin{equation}\label{eq:goreweight}
w = \frac{\sigma_{E}^2(1-\rho\sigma_{L'}/\sigma_{E})}{\sigma_{E}^2 + \sigma_{L'}^2  -2 \rho\sigma_{E}\sigma_{L'}}
\end{equation}
where $\sigma_E^2$ and $\sigma_{L'}^2$ are the variances of $E$ and $L'$, respectively, and $\rho$ is the correlation between them.
The expected variance of  $E^{comb}$ is then:
\begin{equation}
\sigma^2_{E^{comb}} = \frac{\sigma_E^2\sigma_{L'}^2(1-\rho^2)}{\sigma_E^2+\sigma_{L'}^2 -2\rho\sigma_E\sigma_{L'}}\,.
\end{equation}

The application of this algorithm to data improves the energy resolution considerably, 
in particular at high energies where the energy resolution at 2615\un{keV} improves from 16 to 13\un{keV~FWHM}  (see Tab.~\ref{Table:energy resolution} and  Fig.~\ref{fig:decorrelation} right). 

The resolutions of the light detector and of the ZnSe, as well as their correlation, depend on the energy.
To apply Eq.~\ref{eq:goremagari}  to the entire energy spectrum, even where there are no peaks, a specific weight should be used (Eq.~\ref{eq:goreweight}) 
at each energy. Since the resolutions and the correlation are slow functions of the energy, $w(E)$ has been estimated from  polynomial functions
fitting the energy dependence of $\sigma_E$, $\sigma_{L'}$ and $\rho$.

\begin{table}[hbtp]
\centering
\caption{Energy resolution before and after the combination with the detected light.}
\begin{tabular}{lcc}
\hline
                           &ZnSe                     &ZnSe and Light  \\
                           &[keV~FWHM]                                     &[keV~FWHM]      \\
\hline
1461 keV          &13.4 $\pm$ 1.0                  &12.2 $\pm$ 0.8 \\
\hline
2615 keV          &16.3 $\pm$ 1.5                   &13.4 $\pm$ 1.3\\
\hline
\end{tabular}
\label{Table:energy resolution}
\end{table}

\section{Pulse shape discrimination}\label{sec:psa}

Given the different light and heat yields of $\alpha$ and $\beta/\gamma$
interactions, we tested  whether also the shape of the signals carries information
on the type of interacting particle~\cite{Gironi:2009ay,Arnaboldi:2010gj}.

\begin{figure}[t]
\centering
\includegraphics[clip=true,width=0.49\textwidth]{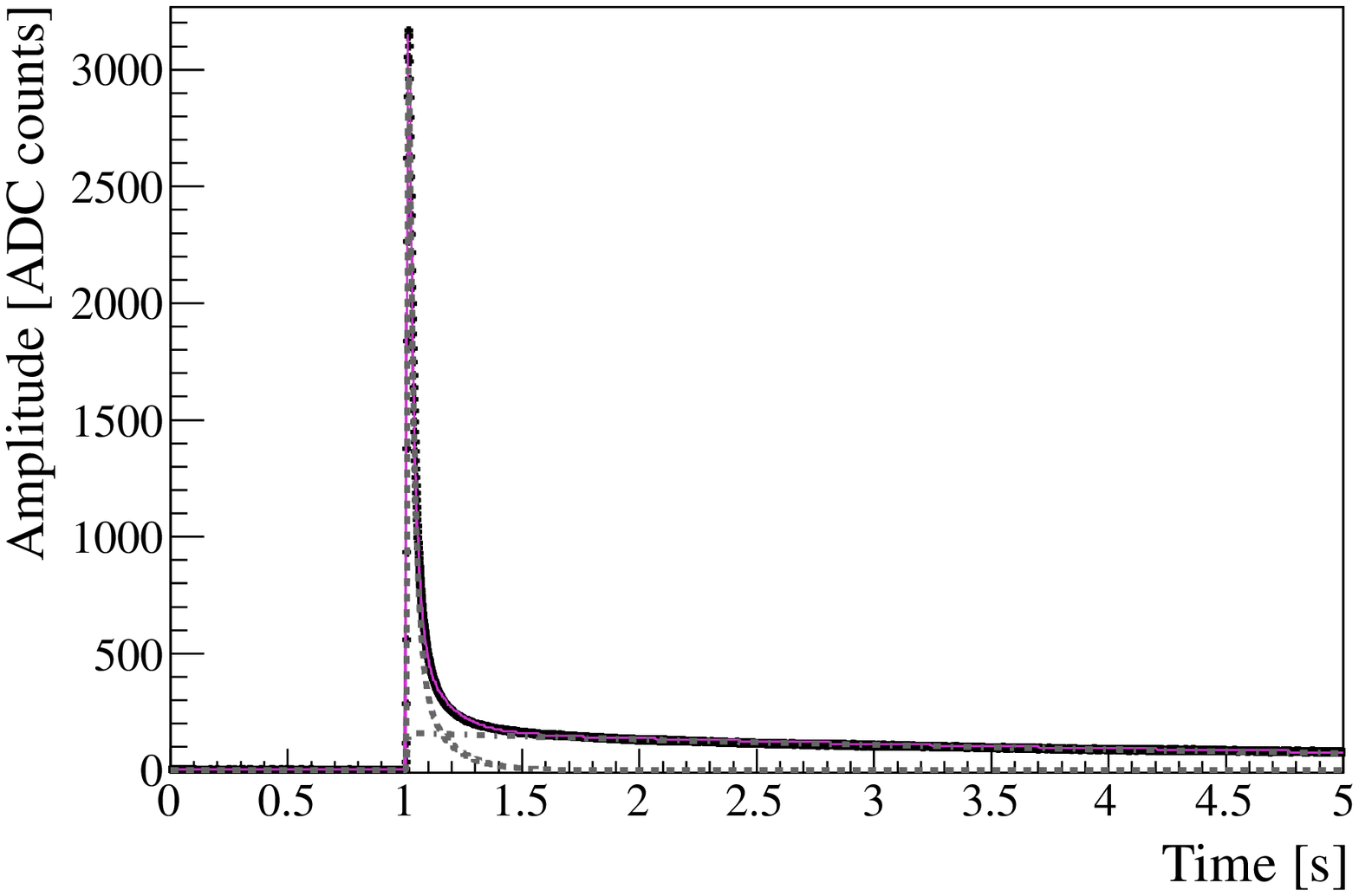}
\hfill
\includegraphics[clip=true,width=0.49\textwidth]{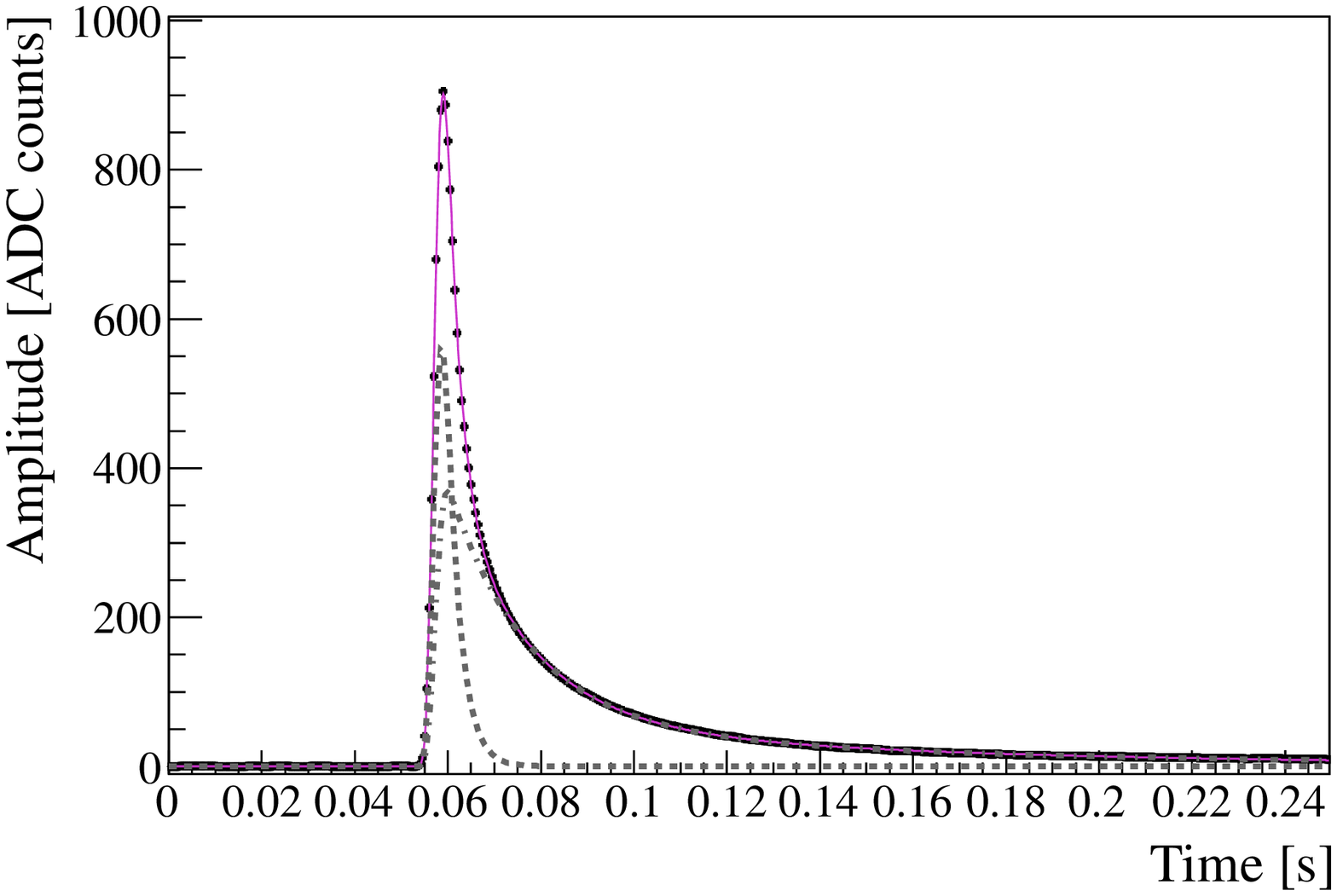}
\caption{
Average pulse of 2615\un{keV} $\gamma$ events in the ZnSe (left)
and the corresponding average scintillation pulse (right). The dashed lines
represent the fast and slow components of the fit function.}
\label{fig:fitpulse} 
\end{figure}

The average pulse of 2615\un{keV} $\gamma$ events in the ZnSe
and the corresponding average scintillation pulse are shown in
Fig.~\ref{fig:fitpulse}.  The pulses are fitted with a model 
developed for \TEO\ bolometers~\cite{Carrettoni:2011rn,Vignati:2010yf}.
The model originally included the thermistor and electronics responses and the
response of the bolometer, which describes the time development of the
thermal phonon signal after an instantaneous energy absorption:
\begin{equation}
\Delta T(t) = A \left[-e^{-t/\tau_r} + \alpha e^{-t/\tau_{d1}} + (1-\alpha) e^{-t/\tau_{d2}}\right]
\end{equation}
where $A$ is the thermal amplitude, $\tau_r$ the rise time, $\tau_{d1,2}$
two decay constants and $\alpha$ is a weight ($0<\alpha<1$). 
In order to introduce the model in this work, the above equation has been modified to include 
energy releases that can not be considered instantaneous, such as the scintillation process.

A function which has been found to well reproduce the observed signals, both in the ZnSe and in the light detector, is:
\begin{equation}
\Delta T(t) = A_1 \left[-e^{-t/\tau_{r}} + \alpha e^{-t/\tau_{d1}} + (1-\alpha) e^{-t/\tau_{d2}}\right] +
 A_2 \left[-e^{-t/\tau_{r}} + e^{-t/\tau_{d3}}\right].
\end{equation}
In Fig.~\ref{fig:fitpulse} the total fit
function and the two components are displayed.
In the ZnSe the amplitude of the fast component is given by $A_1$, while the slow one by $A_2$.
In the light detector the amplitude of the fast component is $A_2$, while the slow one is $A_1$.
We suppose that the fast decay of the signal can be attributed to phonons, 
while the slow one is affected by the scintillation mechanism, which is different for $\alpha$ and $\beta/\gamma$ interactions~\cite{GironiPSD}.
Applying the fit to each event, we noticed that the ratio of the slow and fast amplitude components ($A_s/A_f$)
is related to the particle type.  However the fit is too sensitive to
the detector noise and to minimization problems and cannot be used to estimate a reliable discrimination parameter.

Following Refs.~\cite{Moore:2012au,filippiniphd}, we developed a bi-component optimum
filter algorithm to lower the noise and improve the reliability of
the parameter $A_s/A_f$. To explain the algorithm, we first remind
to the reader the working principles of the standard optimum filter.
Given a signal~+~noise waveform $f(t) = A\cdot S(t-t_0)+ n(t)$, the
best estimate of the amplitude $A$, under the hypothesis that the
noise is stationary, is obtained from the $\chi^2$ of the
residuals in the frequency domain:
\begin{equation}
\chi^2 = \sum_\omega \frac{\left|f(\omega) - A\cdot S(\omega) e^{-i\omega t_0}\right|^2}{N(\omega)}  
\label{eq:chi2}
\end{equation}
where $S(t)$ is the ideal pulse, estimated as the
average pulse,  and $N(\omega)$ is the noise power spectrum of
the detector. The parameter $t_0$ accounts for any possible jitter between
the observed signal and $S(t)$. 
Minimizing Eq.~\ref{eq:chi2} with respect to $A$, we obtain:
\begin{equation}
\hat{A}(t_0) =  h\, \sum_\omega \frac{S^*(\omega)e^{i\omega t_0}}{N(\omega)}f(\omega)
\label{eq:of}
\end{equation}
where $h = [\sum_\omega |S(\omega)|^2/ N(\omega)]^{-1}$.
To estimate $A$, Eq.~\ref{eq:chi2} has to be minimized also with respect to $t_0$.
This however cannot be done in an analytical form.
It can be demonstrated that the minimum of Eq.~\ref{eq:chi2} with respect to $t_0$  and $A$ is equivalent
to the maximum of $\hat A(t_0)$ . Since Eq.~\ref{eq:of} is an inverse Fourier transform on $t_0$,
$\hat{A}(t_0)$ can be seen as the filtered signal in the time domain.
In this domain, the best estimate of $A$ ($\hat{A}$) can be extracted with a maximum search algorithm.

In the bi-component filter the waveform is decomposed as $f(t) = A_s S_s(t-t_0) + A_f S_f(t-t_0) + n(t)$, where in our case $S_s(t)$ and $S_f(t)$
are the ideal slow and fast components of the signal. The minimization of the $\chi^2$ with respect to $A_s$ and $A_f$ gives:
\begin{equation}
 \left(\begin{matrix}
 \hat{A}_{s}(t_0) \\ 
 \hat{A}_{f}(t_0)
 \end{matrix}\right)
 =
 \left(\begin{matrix} 
 \sum \left| {S}_{s}\right|^{2}N^{-1} & \sum ( {S}^{*}_{s} {S}_{f})N^{-1} \\ 
 \sum ( {S}^{*}_{s} {S}_{f})N^{-1}    & \sum \left| {S}_{f}\right|^{2}N^{-1} 
 \end{matrix}\right)^{-1}
 \cdot
 \left(\begin{matrix}
 \sum {S}^{*}_{s}e^{i\omega t_0} N^{-1}\,f\\ 
 \sum {S}^{*}_{f}e^{i\omega t_0}N^{-1}\,f 
 \end{matrix}\right)
\end{equation}
where the sums run over $\omega$ and $S_s,S_f,N$ and $f$ are to be intended in the frequency domain.
In this case the minimum of the $\chi^2$ with respect to $t_0$ does not correspond to the maxima of $\hat{A}_s(t_0)$ and $\hat{A}_f(t_0)$.
Therefore we estimate $\hat{t}_0$ by scanning the $\chi^2$ around $t_0=0$ and then evaluate  $\hat{A}_f=\hat{A}_f(\hat{t}_0)$ and  $\hat{A}_s=\hat{A}_s(\hat{t}_0)$.

\begin{figure}[htbp]
\centering
\includegraphics[clip=true,width=0.49\textwidth]{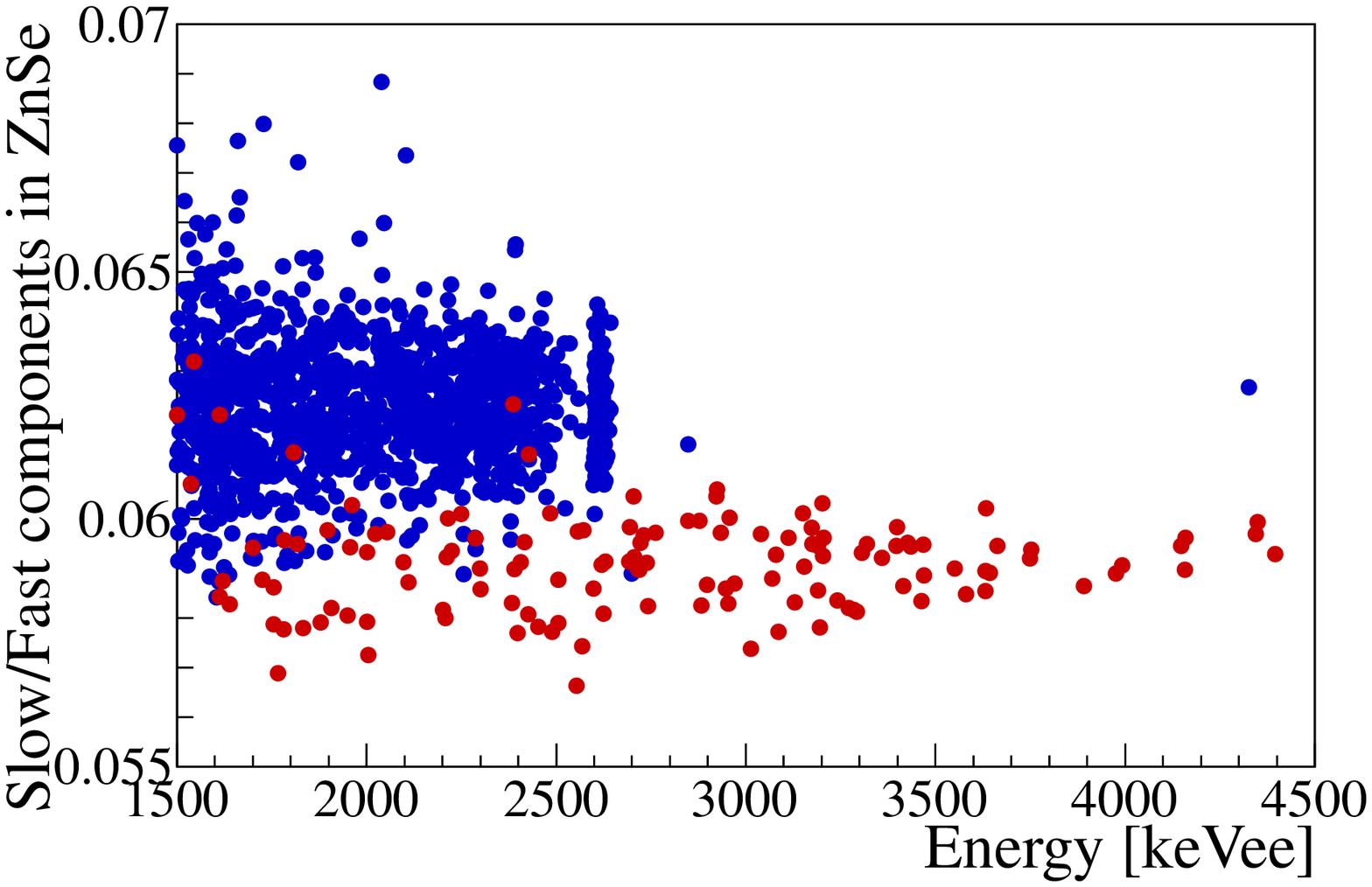}
\includegraphics[clip=true,width=0.49\textwidth]{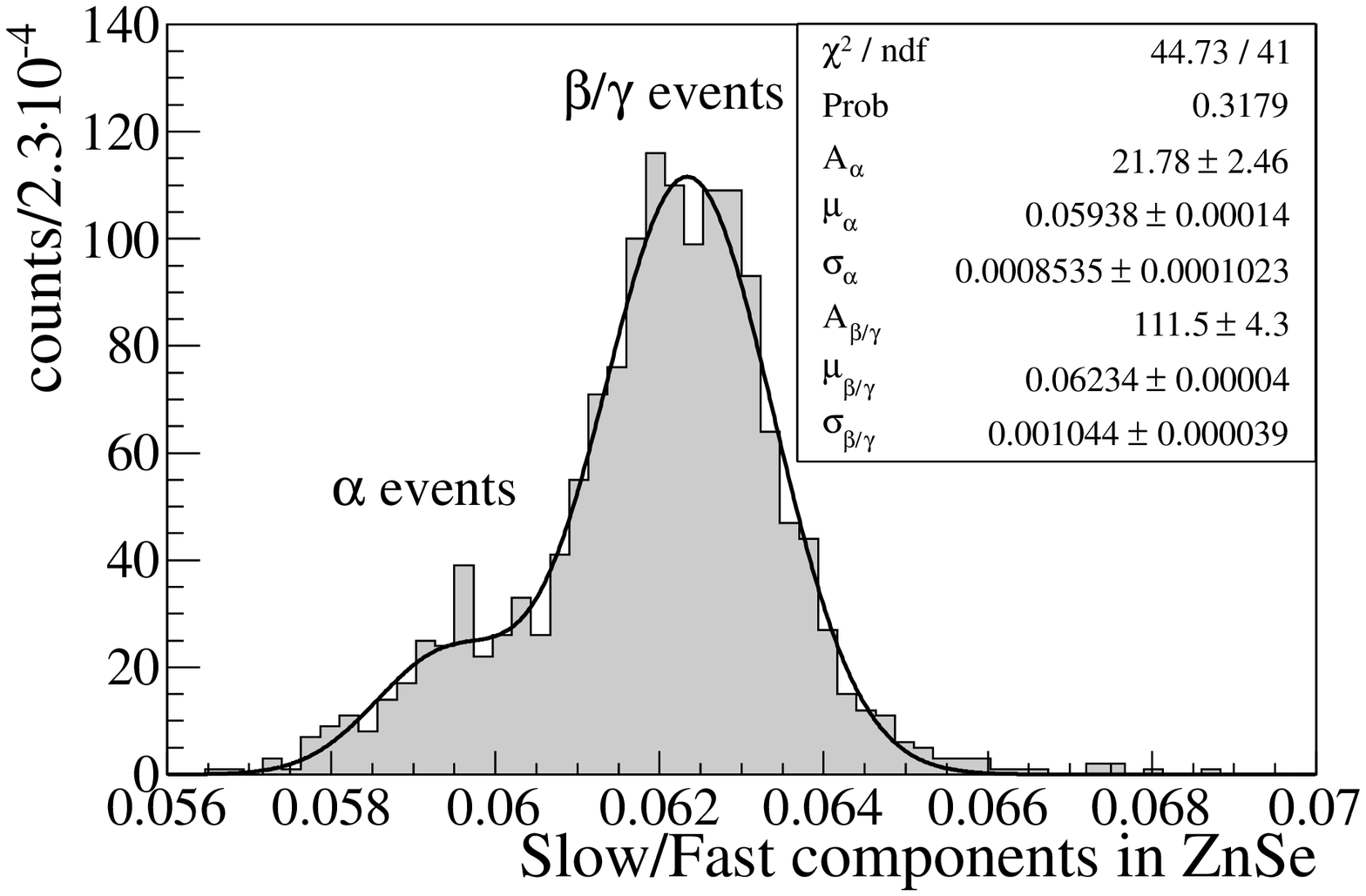}
\includegraphics[clip=true,width=0.49\textwidth]{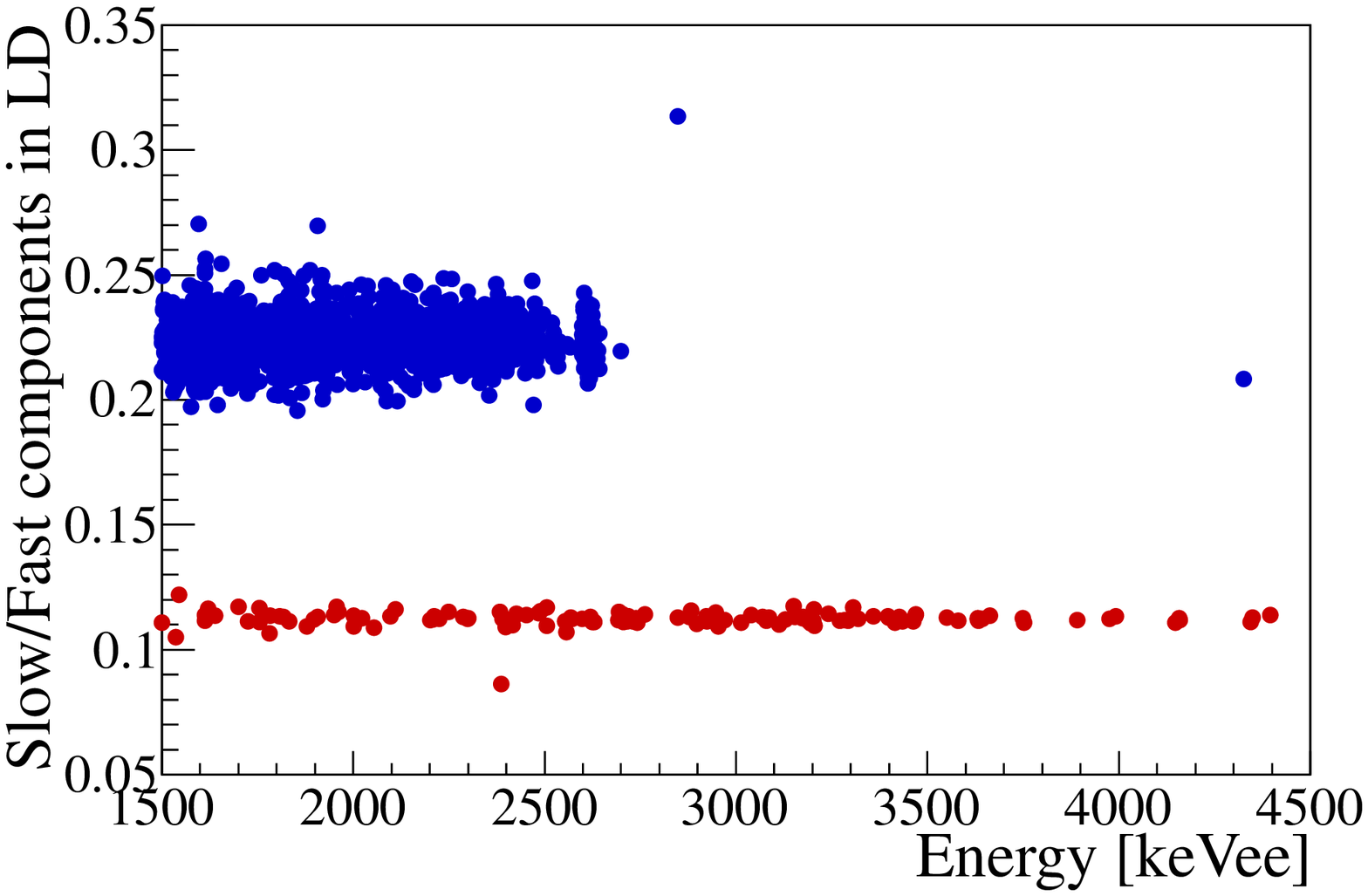}
\includegraphics[clip=true,width=0.49\textwidth]{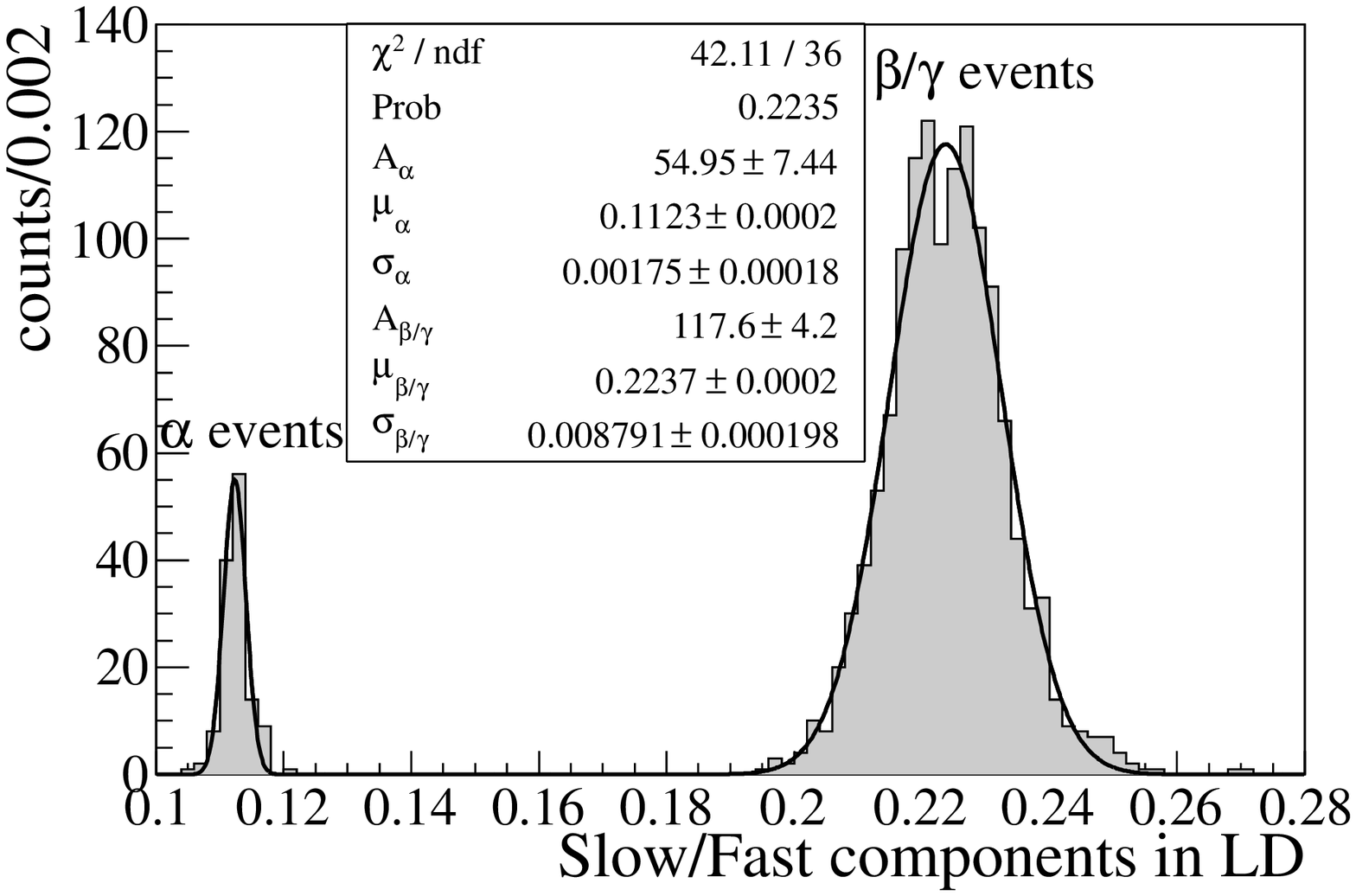}
\hfill
\caption{Left: Ratio of the amplitudes estimated by the bi-component optimum filter algorithm as a function
of the energy released in the ZnSe. The events tagged as $\alpha$ are marked with red, while $\beta/\gamma$s with blue.
Right: Histogram of the ratio. Heat in ZnSe in the top row, light in the LD in the bottom row.  
 The $\alpha$ and $\beta/\gamma$ populations are well identified in the light detector but less in the ZnSe.
}
\label{fig:pulseshapediscr} 
\end{figure}

The application of the bi-component optimum filter is shown in Fig.~\ref{fig:pulseshapediscr},
using the fit components in Fig.~\ref{fig:fitpulse} as estimates of $S_{s,f}(t)$.  As the figure shows, the separation
is very evident in the light detector and less evident in the ZnSe.
In the case of the ZnSe we estimate $DP = 2$, while for the light detector $DP  = 11$.
As it will be shown in the next section, the pulse shape selection applied to the light detector
allows one to tag  $\alpha$ events with light yield compatible  to that of $\beta/\gamma$s, thus dramatically reducing
the background.

\section{Background}

\begin{figure}[tb]
\centering
\includegraphics[clip=true,width=0.7\textwidth]{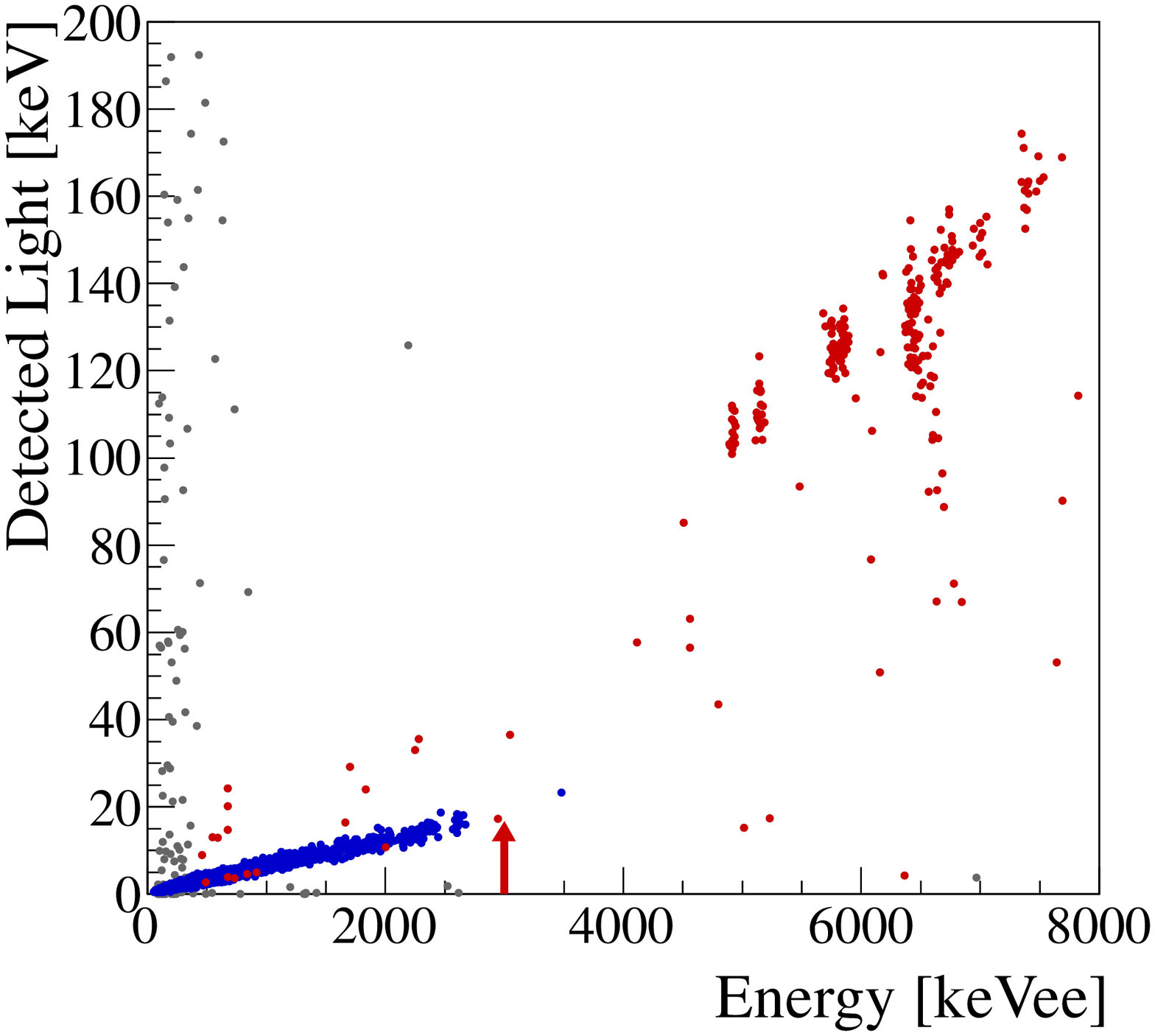}
\caption{Data from  524 hours of background runs. The different colors mark
$\beta/\gamma$ events (blue) and $\alpha$ events (red) selected
with pulse shape cuts on the light channel.
The arrow points to the Q-value of the $^{82}$Se decay. The
gray dots indicate particles interacting in both the ZnSe and the light detector (double hit events), and events with no detected light
(dark events). Double hit events are identified through their pulse shape in the LD, while dark ones
through the shape in the ZnSe.}
\label{fig:light vs heat background}
\end{figure}

We performed a background run of 524 hours in order to evaluate the internal contaminations of the crystal.
To further decrease the environmental $\beta/\gamma$ background, the crystal
was shielded with ancient Roman lead, featuring an activity lower than 4\un{mBq/kg} in $^{210}$Pb~\cite{Alessandrello1998163}.
Figure~\ref{fig:light vs heat background} shows the detected light as a function of the particle energy measured by the ZnSe. 
We notice that the pulse shape cuts allow to completely identify the interacting particle.

\begin{figure}[tb]
\centering
\includegraphics[clip=true,width=0.7\textwidth]{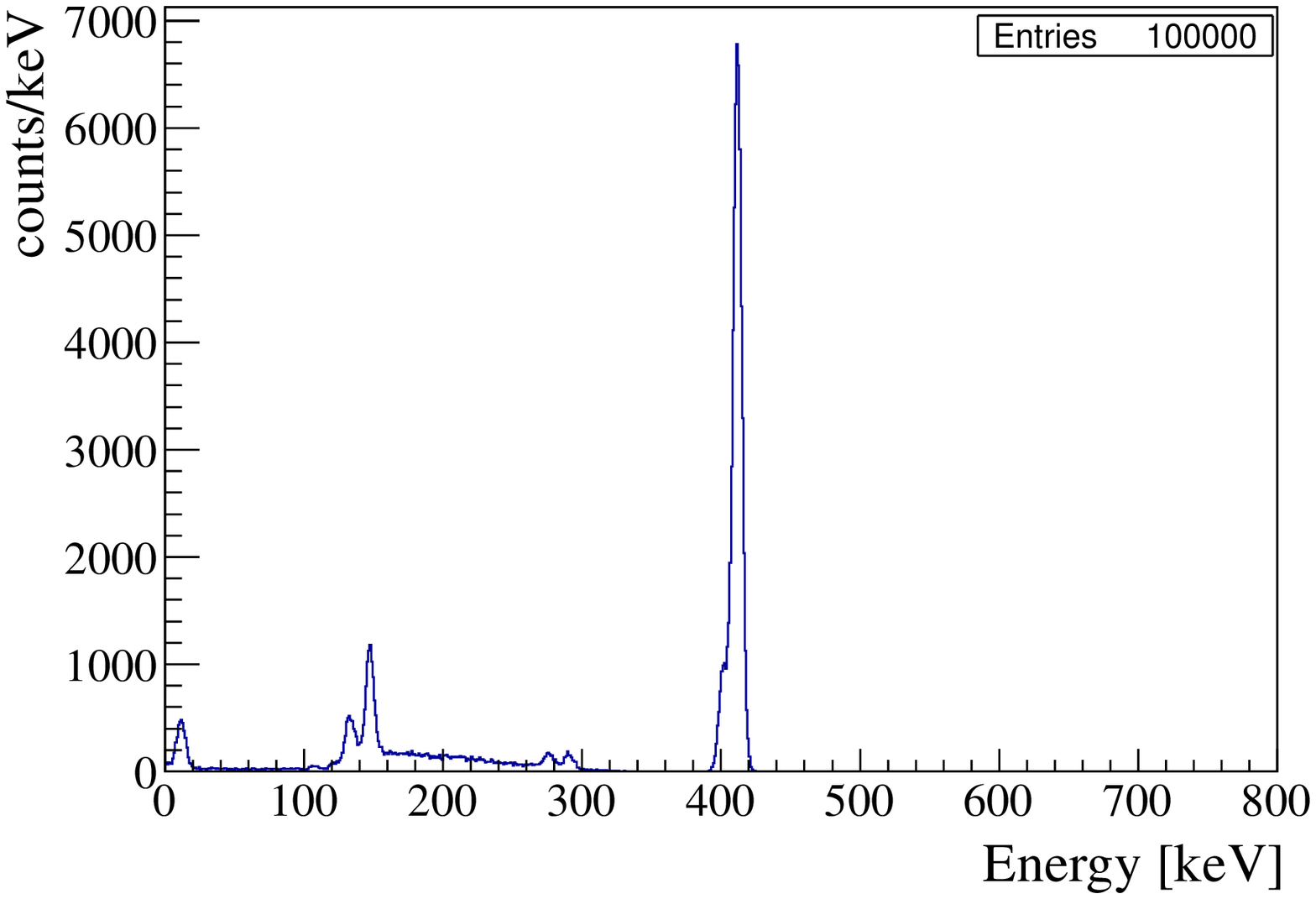}
\hfill
\caption{Simulation of the $^{75}$Se decay in the ZnSe. Energy resolution set to 7\un{keV~FWHM}.}
\label{fig:Simulation 75Se} 
\end{figure}

\begin{figure}[tb]
\centering
\includegraphics[clip=true,width=0.45\textwidth]{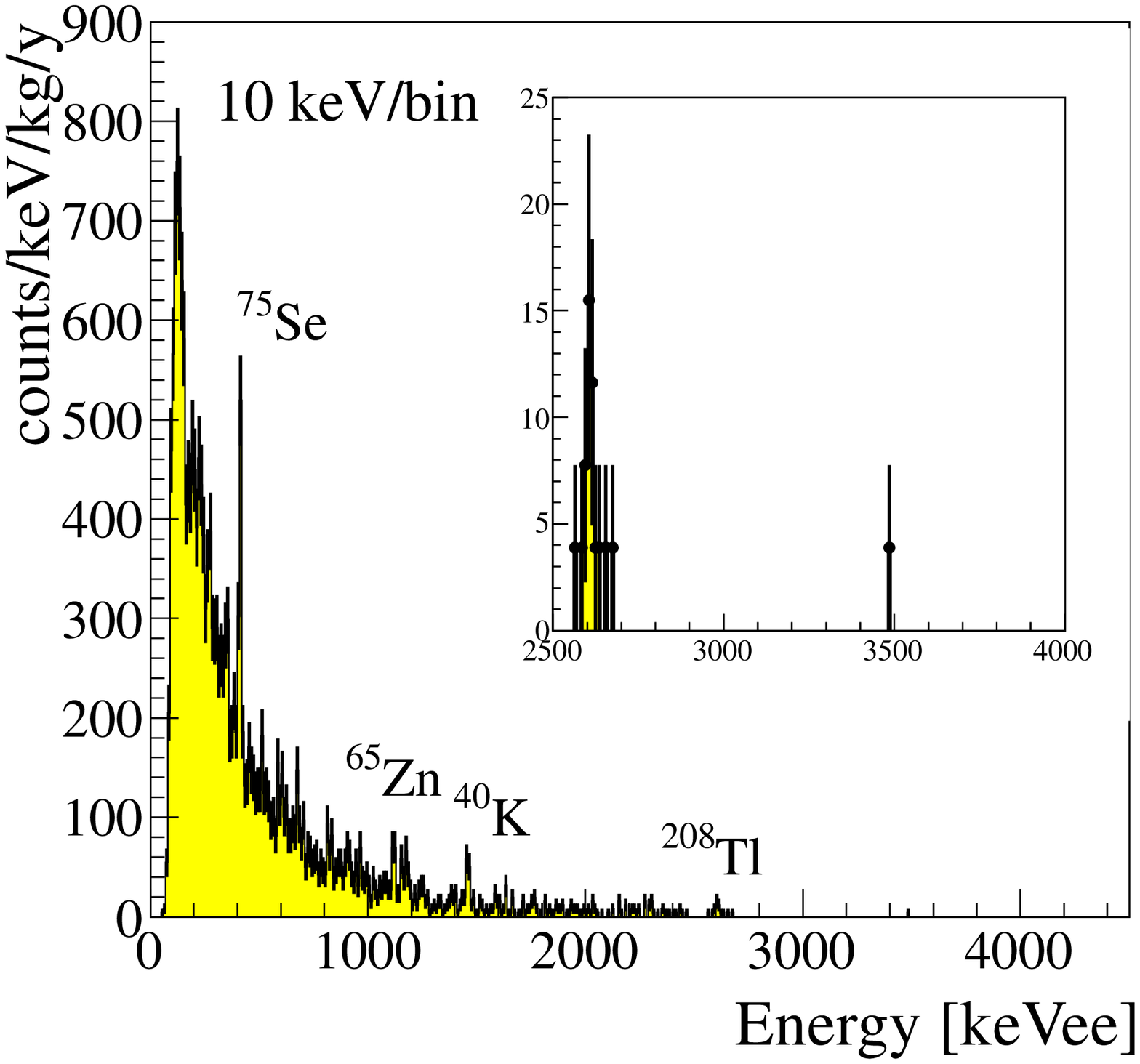}
\hfill
\includegraphics[clip=true,width=0.45\textwidth]{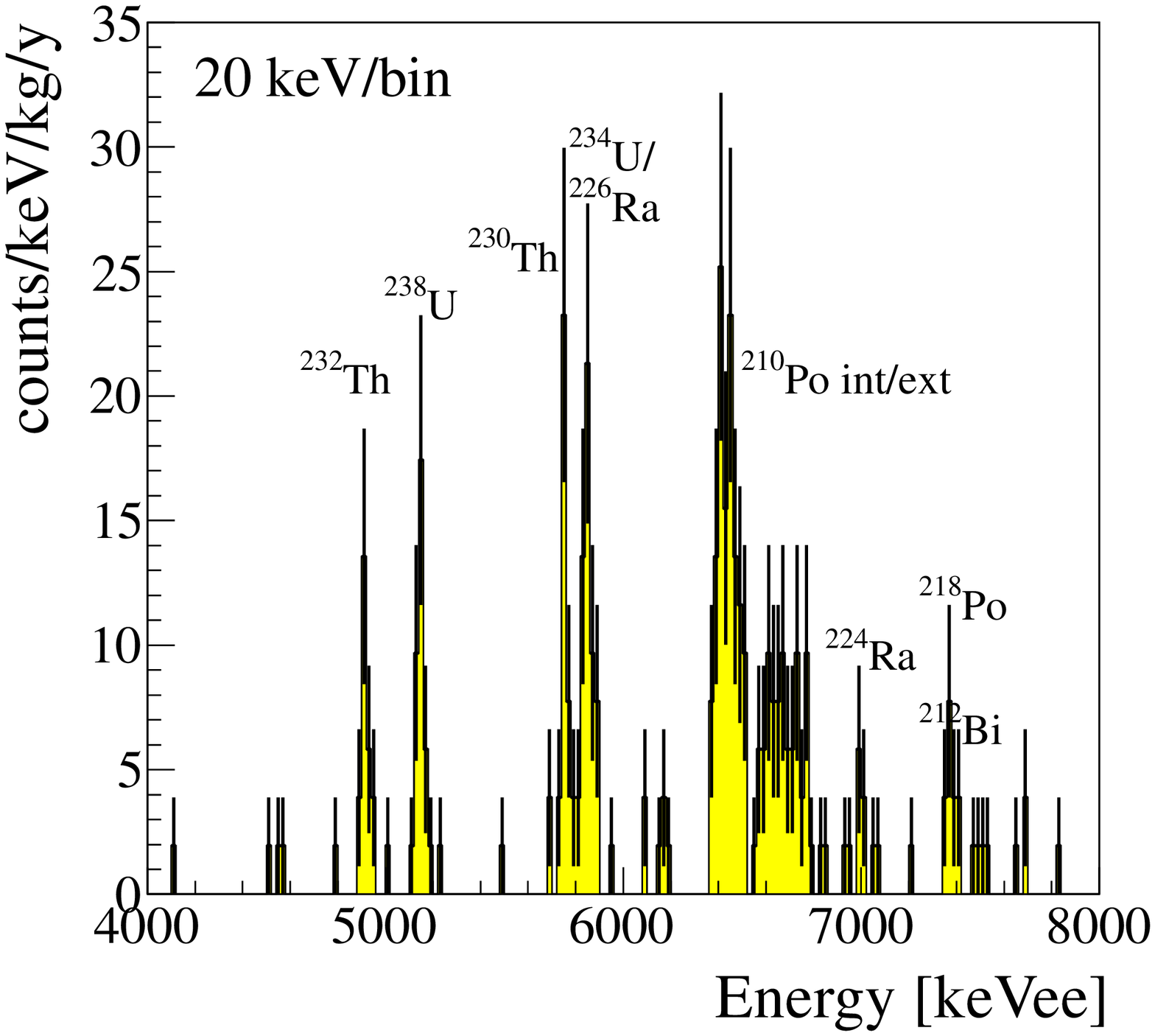}
\hfill
\caption{524\un{h} background run. Spectrum of $\beta/\gamma$ (left) and $\alpha$  events (right). 
Data are calibrated using $\gamma$ sources. $\alpha$s result miscalibrated by +22\% with respect to their nominal energy.}
\label{fig:betaalpha heat spectrum} 
\end{figure}

In Figure \ref{fig:betaalpha heat spectrum} (left) the distribution of $\beta/\gamma$ events is reported. 
We can identify the decay of $^{75}$Se (T$_{1/2} = 119.779\un{days}$, $Q=863.6\un{keV}$).
$^{75}$Se is produced via $^{74}$Se neutron capture and decays via electron-capture (100$\%$) with a complex combination of de-excitation $\gamma$s and X-rays, 
producing a peak at $\sim 410\un{keV}$ and a series of peaks between $150$ and $300\un{keV}$ (Fig.~\ref{fig:Simulation 75Se}).
We analyzed the decay rate of the 410\un{keV} peak as a function of time, obtaining T$_{1/2}$($^{75}$Se) = 102 $\pm$ 18 days,
which confirms our expectations.
Another isotope that is produced by neutron activation in ZnSe is $^{65}$Zn (T$_{1/2} = 244\un{days}$, $Q=1359.9\un{keV}$).
This isotope decays via electron capture and produces a peak at about 1350\un{keV}.
These contaminants do not contribute to the \DBD\ background of $^{82}$Se,
because of the low Q-value and the short half-life.
The other events in the $\beta/\gamma$ spectrum can be attributed to $^{40}$K and $^{232}$Th  contaminations of the environment.

Finally, in the inset of Fig.~\ref{fig:betaalpha heat spectrum} (left), the zoom on the energy region of interest of $^{82}$Se is reported. One can see that only one event above 2615 keV is observed.
This event, however, is in coincidence with high energy $\gamma$'s seen by other scintillating crystals that we were testing in the same set-up. We believe that it can be ascribed to a muon interaction in the surrounding materials.
This kind of events can be easily suppressed in bolometric arrays by applying time-coincidence cuts, as done in CUORICINO~\cite{Andreotti:2010vj} or,
ultimately, by surrounding the cryogenic facility with a muon veto.

In Fig.~\ref{fig:betaalpha heat spectrum} (right) the $\alpha$ background is reported after cutting
on both pulse shape and energy of the light channel. 
The activities of  the most intense peaks, reported in Table \ref{Table: alpha activity}, show that $^{232}$Th and $^{238}$U are in equilibrium with their
daughters. 
\begin{table}[!b]
\centering
\caption{Activity of the isotopes belonging to $^{232}$Th  and $^{238}$U chains.}
\begin{tabular}{lcc}
\hline
Chain              &Nuclide                     &Activity\\
                         &                                   &[$\mu$Bq/kg] \\
\hline
$^{232}$Th    &$^{232}$Th             &17.2 $\pm$ 4.6\\
\hline
                         &$^{228}$Th             &11.1 $\pm$ 3.7\\
\hline
\hline
$^{238}$U    &$^{238}$U                &24.6 $\pm$ 5.5\\
\hline
                       &$^{234}$U                &17.8 $\pm$ 3.3\\
\hline
                       &$^{230}$Th               &24.6 $\pm$ 5.5\\
\hline
                       &$^{226}$Ra              &17.8 $\pm$ 3.3\\
\hline
                       &$^{210}$Po               &$<$ 90.9 $\pm$10.6\\ 
\hline
\end{tabular}
\label{Table: alpha activity}
\end{table}

In principle, a large internal contamination in $^{238}$U could be worrisome because of one of its daughters, $^{214}$Bi, which $\beta$-decays with a Q-value of 3272 keV.
However, $^{214}$Bi decays with a BR of 99.98$\%$ in $^{214}$Po, which $\alpha$-decays with a Q-value of 7.8 MeV. The half-life of  $^{214}$Po, which is about 160 $\mu$s, is extremely short compared to the time development of bolometric signals, which is hundreds of ms. For this reason, the $\beta/\gamma$ emission of $^{214}$Bi is simultaneously followed by a 7.8 MeV $\alpha$ particle and does not represent a problem for the background, 
as the superimposition of the two signals lies at much higher energies than the region of interest.

When evaluating the background due to $^{214}$Bi  we considered also its decay in $^{210}$Tl, which occurs via $\alpha$ emission with a BR of 0.02 $\%$. 
This contribution, however, can be easily suppressed: $^{210}$Tl $\beta$-decays with a Q-value of 5489 keV and a half-life of 1.30 min. 
The background induced by the $\beta$ emission can be rejected using the delayed coincidences with the $\alpha$ of $^{214}$Bi.  
In this case, the dead time induced by the delayed coincidences is almost negligible, as the half-life of $^{210}$Tl is very short and the BR for this decay is extremely small.
As a consequence, even an internal contamination in $^{238}$U of 25 $\mu$Bq/kg would produce a background in the energy region of interest of only 3$\times$10$^{-4}$ counts/keV/kg/y.

A more dangerous background source is the one due to the internal contamination of $^{208}$Tl,
that belongs to the $^{232}$Th chain and $\beta$-decays with a Q-value of 5001 keV.

Considering a $^{232}$Th contamination of 17 $\mu$Bq, we expect a background in the energy region of interest due to $^{208}$Tl of about 2.5$\times$10$^{-3}$ counts/keV/kg/y, 
which represents the ultimate limit for the background. The background due to $^{208}$Tl could be suppressed using the coincidences with its parent, $^{212}$Bi,
that $\alpha$-decays with a half-life of 3 minutes.
In this case, however, the introduced dead-time must be considered. A background reduction factor of 3, for example, requires a delayed coincidence of 10\un{min}, which would imply a dead time of about 10$\%$.

Summarizing, the internal contaminations of $^{214}$Bi and $^{208}$Tl do not represent a dangerous problem for the achievement of a background of 10$^{-3}$ counts/keV/kg/y, 
which is the goal of the LUCIFER experiment. 
In any case, purification techniques are under study in order to reduce bulk contaminations of the crystals and, therefore, the dead time induced by delayed coincidences.

\section{Conclusions}

In this paper we analyzed the performances of a 431\un{g} ZnSe crystal
operated as scintillating bolometer.  We measured the energy resolution
at 2615\un{keV}, close to the \DBD\ of $^{82}$Se (2997\un{keV}).
By combining the light and heat signals, the resolution has been estimated
as 13\un{keV~FWHM}.

We presented the results of a 524 hours background run to identify
the internal radioactive contaminations of the crystal.  We measured $^{232}$Th and $^{238}$U
contaminations of the order of tens of $\mu$Bq/kg, which is compatible with the low background requirements of a \DBD\ experiment.  
In this run, no event passed the data analysis cuts
against the background in the energy region of the decay,
showing the potential of this detection technique.

We analyzed the scintillation properties of the detector and we
demonstrated that the information carried by the light signal allows to
identify the nature of the interacting particles in two different ways.
Exploiting the different light yield of $\alpha$s and
$\beta/\gamma$s, one can reject the background due to $\alpha$ particles
within the energy region of interest for the $^{82}$Se decay.
We developed an algorithm,
based on the shape of the light signals, to tag $\alpha$ particles
with light yield compatible with $\beta/\gamma$ ones. 

Finally, the good identification of nuclear
recoils at 100\un{keV}  allowed us to set the requirements for Dark Matter searches
with ZnSe bolometric detectors. We estimated that nuclear recoils of $^{210}$Po are detected with
an amount of energy 35\% greater than the nominal value, using a calibration function
estimated from $\gamma$ lines. This value has to be confirmed with dedicated tests,
together with the evaluation of the calibration function of  Zn and Se recoils.
By using light detectors with baseline noise lower than 20\un{eV~RMS},
we estimate that $\beta/\gamma$ events could be discriminated from nuclear recoils events of $10\un{keV}$,
thus setting indicative requirements for Dark Matter searches in LUCIFER.

\acknowledgments
This work was partially supported by the LUCIFER experiment, funded by ERC under the European Union's Seventh Framework Programme (FP7/2007-2013)/ERC grant agreement n. 247115, funded within the ASPERA 2nd Common Call for R\&D Activities.
Thanks are due to the LNGS mechanical workshop and in particular to E. Tatananni, A. Rotilio, A. Corsi, and B. Romualdi for continuous and constructive help in the overall set-up construction. Finally, we are especially grateful to M. Guetti for his help.

\bibliographystyle{JHEP} 
\bibliography{main}

\end{document}